\documentclass[%
 reprint,
 superscriptaddress,
 amsmath,amssymb,
 aps,
 pre,
 floatfix,
]{revtex4-2}

\usepackage{graphicx}
\usepackage[caption=false]{subfig}
\usepackage{dcolumn}% Align table columns on decimal point
\usepackage{bm}% bold math
\usepackage{wrapfig}
\usepackage{enumitem}
\usepackage{amsmath,stackengine}
\stackMath
\usepackage{amssymb}
\usepackage{marginnote}
\usepackage{comment}
\usepackage{url}
\usepackage{esint}
\usepackage{multirow}
\usepackage[caption=false]{subfig}
\usepackage{tikz}
\usetikzlibrary{arrows.meta}

\DeclareMathOperator{\Arg}{Arg}

\usepackage{mathtools}
\usepackage{listings}
\usepackage[dvipsnames]{xcolor}
\usepackage[colorlinks, allcolors=NavyBlue]{hyperref}
\usepackage{xspace}
\usepackage{mathrsfs}
\usepackage{physics}
\usepackage[normalem]{ulem}

\begin{document}

\title{Active-learning mapping of the Vicsek model phase diagram}

\author{Grace T. Bai}
\affiliation{Department of Computer Science, University of Virginia, Charlottesville, Virginia 22904, USA}
\author{Brandon B. Le}
\affiliation{
Department of Physics, University of Virginia, Charlottesville, Virginia 22904, USA
}

\date{\today}

\begin{abstract}
The Vicsek model is a minimal model of collective motion, capturing how local alignment interactions can generate macroscopic nonequilibrium order in systems such as bird flocks. In this work, we use active learning to map the Vicsek phase diagram as a function of noise strength, density, and particle speed. A neural-network classifier is trained on global polar-order labels, and classifier entropy is used to select new simulations near uncertain crossover regions. The resulting phase map resolves a high-noise disordered gas, a low-noise polar ordered regime, and an intermediate coexistence-candidate regime whose noise window shifts upward and broadens with increasing density. Independent density and local-order diagnostics indicate that the intermediate regime contains dense, locally ordered bands coexisting with a dilute, weakly ordered background. Comparison with the ordered regime shows that banded coexistence is identified by the joint enhancement of band contrast, local-order heterogeneity, and positive density-order correlation. Overall, these results establish a machine-learning-guided workflow for active matter, in which active learning constructs an operational phase map and independent spatial diagnostics convert classifier-defined regimes into physically interpretable nonequilibrium morphologies.
\end{abstract}

\maketitle

%\tableofcontents

\color{black}

\section{Introduction}

Collective motion is one of the central examples of emergent order in nonequilibrium physics. Systems composed of many individually driven units can spontaneously organize into coherent large-scale motion through local interactions alone. This phenomenon appears across many scales, including bird flocks~\cite{emlen1952flocking, bialek2012statistical}, bacterial colonies and suspensions~\cite{wu2000particle}, cytoskeletal materials~\cite{needleman2017active}, epithelial cell layers~\cite{angelini2011glass}, pedestrian crowds~\cite{warren2018collective}, and robotic swarms~\cite{brambilla2013swarm, rubenstein2014programmable}. Such systems are now understood as active matter, where each constituent consumes energy locally and uses it to generate motion or mechanical stress, placing the many-body system intrinsically far from thermal equilibrium \cite{ramaswamy2010mechanics, marchetti2013hydrodynamics, bechinger2016active, gompper2020motile}. The study of collective motion therefore connects biological and engineered
swarms to broader questions in nonequilibrium statistical physics, including how local self-propelled interactions generate macroscopic order and how phase transitions are organized in systems driven far from equilibrium.

The Vicsek model provides a minimal setting in which to study the emergence of collective motion from local alignment~\cite{vicsek1995novel}. In its standard angular-noise form, point particles move at fixed speed and update their headings by aligning with nearby neighbors, subject to random angular perturbations. This simple rule produces a nonequilibrium transition from a high-noise disordered gas to a low-noise polar flock, but the structure of that transition is considerably richer than the original minimal formulation might suggest. Early work interpreted the onset of collective motion as a continuous transition~\cite{vicsek1995novel}, whereas later numerical studies found strong finite-size effects, discontinuous signatures, and traveling high-density ordered bands near the onset~\cite{gregoire2004onset, chate2008collective, baglietto2012criticality}. Hydrodynamic theories of polar active matter explained how two-dimensional flocks circumvent equilibrium constraints on long-range order and predicted the strong coupling between density and orientation fluctuations~\cite{toner1995long, toner1998flocks, mahault2019quantitative}. The banded regime has also been incorporated into a nonequilibrium liquid-gas or microphase-separation picture~\cite{solon2015phase, ginelli2016physics, chate2020dry}. Recent work has further expanded the phenomenology of Vicsek-type systems, including the discovery of a self-organized cross-sea phase in very large dry active systems~\cite{kursten2020dry}, identification of cluster phases in the Vicsek model~\cite{miyahara2025vicsek}, correlation-length diagnostics for distinguishing transition scenarios~\cite{yu2025unifying}, and modified interaction rules, including random interaction connections and delayed alignment, that reshape band formation and phase separation~\cite{jin2025random, horton2025order}.

For the purposes of the present finite-size study, the relevant phenomenology can be organized into three regimes. At high angular noise, alignment is overwhelmed and the system behaves as a disordered gas, with weakly correlated headings and small global polar order. At low noise, local alignment produces a polar ordered flock, or ordered
liquid in the nonequilibrium liquid-gas interpretation, with a persistent macroscopic direction of motion. Between these limits, the angular-noise Vicsek model can display banded coexistence, in which dense, locally ordered traveling bands are embedded in a dilute, weakly ordered background~\cite{gregoire2004onset, chate2008collective, solon2015phase, chate2020dry}. Resolving this intermediate regime requires more than a global order parameter, since its defining feature is spatial coexistence between locally ordered and weakly ordered regions.

Mapping these regimes over parameter space is computationally demanding. The standard Vicsek model is controlled by the angular noise $\eta$, the density $\rho$, and the particle speed $v_0$, and the relevant crossover surfaces can occupy narrow, curved regions of this three-dimensional space. A uniform grid therefore spends many simulations deep inside already identifiable regimes while still requiring high resolution near the crossovers. This motivates an adaptive strategy in which new simulations are selected according to the uncertainty of the current phase map~\cite{settles2009active, lookman2019active}. Machine-learning methods have become widely used for identifying phases and phase transitions in many-body systems~\cite{carrasquilla2017machine, van2017learning, carleo2019machine, carrasquilla2021neural}, and active learning has been developed as a way to sample phase diagrams efficiently by placing new simulations near uncertain or informative regions~\cite{dai2020efficient, zhu2024active}. Related data-driven approaches have also been applied to collective-motion and active-matter systems~\cite{bhaskar2019analyzing, xue2023machine, miyahara2025vicsek}. These developments suggest a natural strategy for the Vicsek problem: use machine learning to guide sampling of the finite-size phase map, then use physical observables to interpret the learned regimes.

Following this strategy, we use active learning to construct a finite-size phase map of the two-dimensional angular-noise Vicsek model in the control-parameter space $\left(\eta,\rho,v_0\right)$. Simulated parameter points are assigned operational labels from the steady-state global polar order $\left\langle\Phi\right\rangle$, and a neural-network classifier is trained to interpolate between disordered, coexistence-candidate, and polar ordered regimes. The classifier uncertainty is then used to select additional simulations near the learned crossover surfaces, allowing the finite-size phase geometry to be resolved with a compact data set. The learned map is used both to quantify the crossover structure and to guide physical validation. From a fixed-speed slice, we extract effective ordered-to-coexistence and coexistence-to-disorder boundary curves and examine how the coexistence-candidate window changes with density. The learned map then serves as a guide for a second stage of analysis, in which representative intermediate states are examined using spatially resolved density and local-order measurements.

The remainder of the paper is organized as follows. In Sec.~\ref{sec:model_method}, we define the Vicsek model, simulation protocol, cell-list implementation, and observables used throughout the work. In Sec.~\ref{sec:active_learning}, we describe how simulated parameter points are assigned operational labels and how classifier entropy is used to adaptively select new simulations. In Sec.~\ref{sec:phase_diagram}, we use the trained classifier to construct the learned phase map and analyze its probability structure, entropy field, and effective crossover curves. In Sec.~\ref{sec:spatial_validation}, we test the physical content of the intermediate regime using spatial diagnostics and show that it corresponds to banded coexistence along the validated cut. Finally, in Sec.~\ref{sec:conclusion}, we summarize the main results and discuss how this framework can be extended to other questions in nonequilibrium flocking.

\section{Model and Numerical Methods}
\label{sec:model_method}

We consider the standard two-dimensional Vicsek model with angular noise and periodic boundary conditions~\cite{vicsek1995novel}. The system consists of $N$ point particles in a square domain of side length $L$, so the particle density is $\rho = N/L^2$. Particle $i$ has position
\begin{equation}
    \mathbf{r}_i(t) = \left(x_i(t),y_i(t)\right)
\end{equation}
and heading angle $\theta_i(t)$. Its corresponding unit velocity direction is
\begin{equation}
    \mathbf{e}_i(t) = \left(\cos\theta_i(t), \sin\theta_i(t)\right).
\end{equation}
At each time step, particle $i$ aligns its heading with the average direction
of all particles within an interaction neighborhood given by
\begin{equation}
    \mathcal{N}_i(t) = \left\{j: \left|\mathbf{r}_j(t)-\mathbf{r}_i(t)\right|_{\rm pbc} < r_0\right\},
\end{equation}
where $\left|\cdot\right|_{\rm pbc}$ denotes the minimum distance under
periodic boundary conditions. The heading update is
\begin{equation}
    \theta_i(t+\Delta t) = \Arg\left[\sum_{j\in\mathcal{N}_i(t)} e^{i\theta_j(t)}\right] + \eta\xi_i(t),
    \label{eq:vicsek_heading_update}
\end{equation}
where $\eta\in[0,1]$ is the angular noise strength and $\xi_i(t)$ is chosen from a uniform distribution $\xi_i(t) \sim {\rm Unif}[-\pi,\pi]$. The position is then updated using the new heading:
\begin{equation}
    \mathbf{r}_i(t+\Delta t) = \mathbf{r}_i(t) + v_0\,\mathbf{e}_i(t + \Delta t)\,\Delta t \pmod L,
    \label{eq:vicsek_position_update}
\end{equation}
where $v_0$ is the self-propulsion speed. The speed $v_0$ is retained as an explicit control parameter because it affects the transport of orientational information through the system. In the small-speed limit, particles remain in nearly fixed local neighborhoods over many alignment updates, while larger $v_0$ increases spatial mixing and can shift the finite-size crossover locations~\cite{vicsek1995novel, baglietto2012criticality}. Equations~\eqref{eq:vicsek_heading_update} and \eqref{eq:vicsek_position_update} define the discrete-time stochastic update rule of the Vicsek model. Throughout this paper, the interaction radius and time step are fixed to $r_0 = 1$ and $\Delta t = 1$.

A direct implementation of the neighborhood search would require $O(N^2)$ pairwise distance checks per time step. To simulate high-density systems efficiently, we instead use a spatial cell-list algorithm~\cite{allen2017computer}. The domain is divided into a grid of square cells with cell width $w\geq r_0$. Each particle is assigned to a cell, and a linked list stores the particles belonging to each cell. For each particle, candidate neighbors are searched only in the particle's own cell and the eight adjacent cells, with periodic wrapping of cell indices. At fixed density, this reduces the expected neighbor-search cost from $O(N^2)$ to approximately $O(N)$ per time step.

The active-learning phase map is constructed at fixed box size $L = 128$ while varying the parameter vector
\begin{equation}
    \boldsymbol{\lambda} = \left(\eta,\rho,v_0\right).
\end{equation}
The sampled parameter ranges are
\begin{equation}
    \eta\in[0,1], \quad\rho\in[0.1,3.0], \quad v_0\in[0.01,0.5],
\end{equation}
and for each sampled density, the particle number is chosen as $N(\rho) = \operatorname{round}\left(\rho L^2\right)$. 

Throughout each simulation, we record the global polar order parameter, which is defined as
\begin{equation}
    \Phi(t) = \frac{1}{N}\left|\sum_{j=1}^{N}e^{i\theta_j(t)}\right|.
    \label{eq:global_order_parameter}
\end{equation}
For each simulated parameter point, we discard an initial transient and compute steady-state averages over the remaining recorded trajectory. If $\mathcal{W}_{\rm ss}$ denotes the set of recorded times in the steady-state window, then
\begin{equation}
    \left\langle A \right\rangle = \frac{1}{\left|\mathcal{W}_{\rm ss}\right|} \sum_{t\in\mathcal{W}_{\rm ss}} A(t).
\end{equation}
In particular, the steady-state mean order parameter is
\begin{equation}
    \left\langle \Phi \right\rangle = \frac{1}{\left|\mathcal{W}_{\rm ss}\right|} \sum_{t\in\mathcal{W}_{\rm ss}} \Phi(t).
\end{equation}
We also compute the susceptibility-like fluctuation measure
\begin{equation}
    \chi = N\left(\left\langle \Phi^2 \right\rangle - \left\langle \Phi \right\rangle^2\right),
    \label{eq:susceptibility}
\end{equation}
and the Binder cumulant~\cite{binder1981finite}
\begin{equation}
    G = 1 - \frac{\left\langle \Phi^4 \right\rangle}{3\left\langle \Phi^2 \right\rangle^2}.
    \label{eq:binder_cumulant}
\end{equation}
The order parameter $\left\langle\Phi\right\rangle$ is used for operational phase labeling, while $\chi$ and $G$ provide additional diagnostics of fluctuations and non-Gaussian order-parameter statistics near crossover regions.

We use two simulation protocols. For the active-learning phase-map construction, each parameter point $\boldsymbol{\lambda} = \left(\eta,\rho,v_0\right)$ is simulated using three independent random initial conditions, and the resulting steady-state observables are averaged over trials. For the detailed physical validation at fixed density and speed, we use $\rho_0=2.0,v_0=0.16$ and run five independent trials for each representative noise value. 

The numerical kernels for cell-list construction, heading updates, position updates, and order-parameter evaluation are compiled using Numba, and parallelism over parameter points and independent trials is handled at the process level. This implementation makes it feasible to perform repeated high-density simulations while retaining enough independent trials to estimate steady-state fluctuations.

\begin{figure*}[tbp!]
    \centering
    \includegraphics[width=0.9\linewidth]{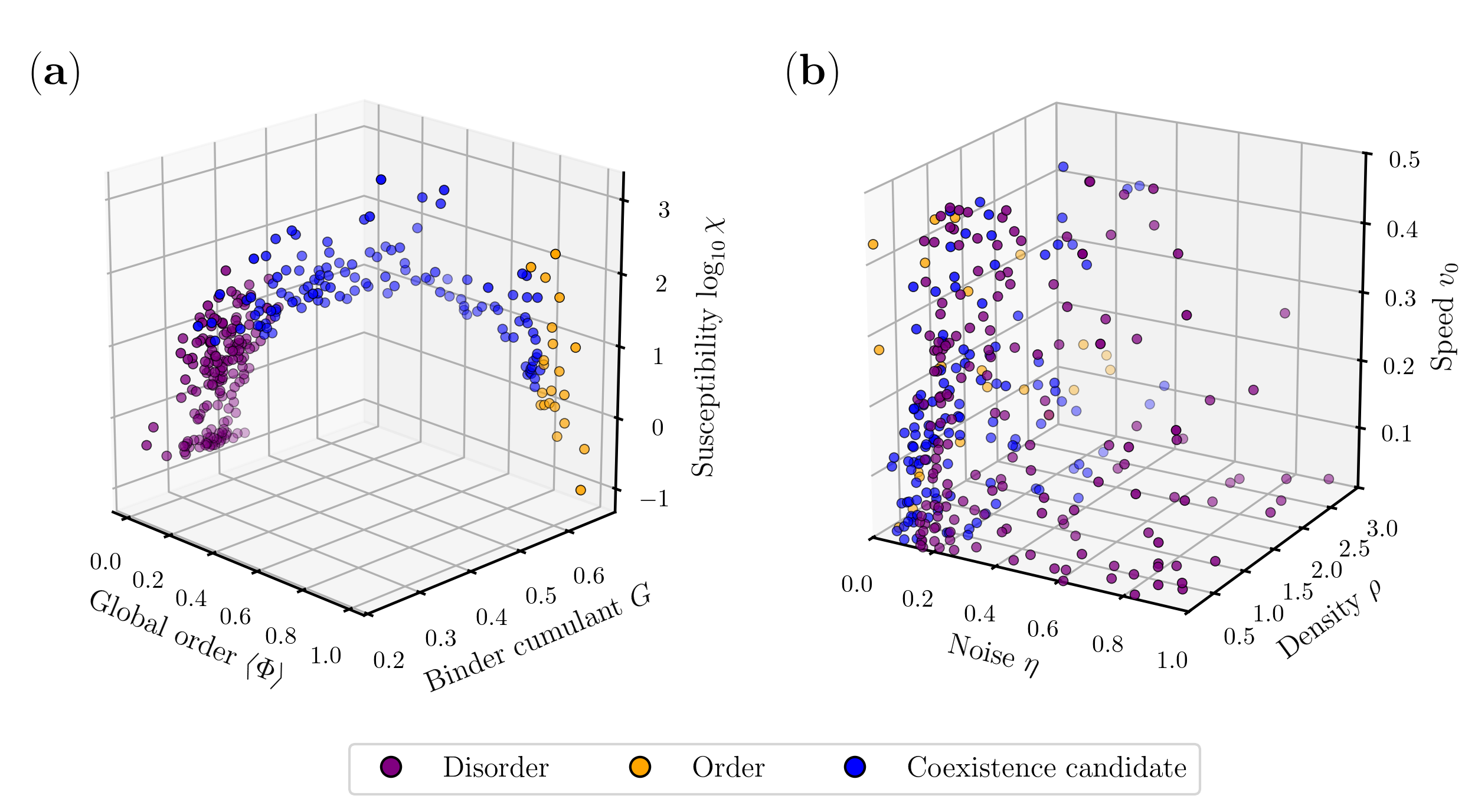}
    \caption{Active-learning data set used to train the phase classifier. Each point represents a simulated parameter configuration, with steady-state observables averaged over independent trials and colored by the operational labels defined from $\left\langle\Phi\right\rangle$. (a)~The data shown in observable space, $\left(\left\langle\Phi\right\rangle,G,\log_{10}\chi\right)$. The disordered states cluster near small global polar order, while the ordered states occupy the large-$\left\langle\Phi\right\rangle$ region. The coexistence-candidate points lie between these limits and are associated with enhanced susceptibility, reflecting the large order-parameter fluctuations typical of crossover states. (b)~The data shown in control-parameter space, $\left(\eta,\rho,v_0\right)$. The distribution combines the initial space-filling Latin-hypercube sample with additional active-learning points selected by maximizing the classifier Shannon entropy, which preferentially populate the regions separating the operational classes.}
    \label{fig:al_dataset}
\end{figure*}

\section{Active-Learning Phase Classification}
\label{sec:active_learning}

The simulations described in Sec.~\ref{sec:model_method} define a map from control parameters $\boldsymbol{\lambda} = \left(\eta,\rho,v_0\right)$ to the steady-state observables $\left\langle\Phi\right\rangle$, $\chi$, and $G$. A dense grid in this three-dimensional parameter space would be inefficient, since most points lie deep inside already identifiable regimes. We therefore use an active-learning procedure to concentrate new simulations near the crossover regions~\cite{settles2009active}. The classifier is then used as both a sampling and interpolation tool, providing a smooth operational phase map and identifying uncertain points for additional simulation.

Each simulated parameter point is assigned an operational label using the steady-state mean polar order parameter $\left\langle\Phi\right\rangle$ defined in Eq.~\eqref{eq:global_order_parameter}. These labels provide a reproducible order-parameter-based rule for training the classifier and for separating clearly disordered, clearly ordered, and intermediate finite-size states:
\begin{align}
    \left\langle\Phi\right\rangle < 0.12 \qquad &\Longrightarrow \quad \text{disordered gas}, \label{eq:label_disorder} \\
    0.12 \leq \left\langle\Phi\right\rangle \leq 0.80 \quad &\Longrightarrow \quad \text{coexistence candidate}, \label{eq:label_coexistence} \\
    \left\langle\Phi\right\rangle > 0.80 \qquad &\Longrightarrow \quad \text{polar ordered regime}. \label{eq:label_order}
\end{align}
We use fixed order-parameter thresholds rather than clustering-based labels so that the training labels remain independent of the changing sampling density produced by active learning. The lower threshold is chosen to be well separated from finite-size polar order generated by random headings.  For a completely disordered configuration with independent headings, the expected order-parameter scale is $\Phi_{\rm noise}\sim 1/\sqrt{N}$. Then, even at the lowest density in the sampled range, $\rho=0.1$, the particle number is $N\approx 0.1L^2 = 1638$, giving $\Phi_{\rm noise}\simeq 0.025$, well below the cutoff $\left\langle\Phi\right\rangle=0.12$. States below this threshold are therefore safely classified as globally disordered rather than weakly ordered finite-size fluctuations. The upper threshold $\left\langle\Phi\right\rangle=0.80$ selects states with strong macroscopic polar alignment and excludes partially ordered or strongly heterogeneous configurations from the ordered class. The remaining interval is then assigned to an intermediate class. However, an intermediate value of $\left\langle\Phi\right\rangle$ alone does not determine the spatial morphology of the state; in particular, it does not by itself distinguish a traveling band from a homogeneous state with reduced global order or large density fluctuations. For this reason, we refer to the intermediate label as a coexistence candidate at this classification stage. Its identification with banded coexistence is tested independently in Sec.~\ref{sec:spatial_validation} using density contrast, density-order correlation, and local polar-order variation.

The susceptibility $\chi$ and Binder cumulant $G$ defined in Eqs.~\eqref{eq:susceptibility} and \eqref{eq:binder_cumulant} are not used as threshold variables for the final labels but still provide diagnostic information about the steady-state fluctuations of $\Phi(t)$. Large values of $\chi$ indicate parameter regions where the global order parameter varies strongly over time or between trials. The Binder cumulant is sensitive to the shape of the order-parameter distribution and is useful for identifying broad or non-single-peaked distributions near crossover regions. This distinction is useful because the active-learning labels are deliberately simple, while the physical classification is checked using additional observables.

The initial training set consists of $100$ parameter points generated by Latin hypercube sampling~\cite{mckay1979comparison}. The noise coordinate is sampled uniformly, while density and speed are sampled uniformly on logarithmic scales:
\begin{align}
    \log_{10}\rho &\sim {\rm Unif}\left(\log_{10}0.1,\log_{10}3.0\right),
    \label{eq:rho_lhs} \\
    \log_{10}v_0 &\sim {\rm Unif}\left(\log_{10}0.01,\log_{10}0.5\right).
    \label{eq:v0_lhs}
\end{align}
The logarithmic sampling is used because both parameters span broad dynamical ranges and control alignment through encounter and transport rates. Each sampled point is simulated using three independent random initial conditions, and the steady-state observables are averaged over trials before assigning the operational label.

We train a feed-forward neural-network classifier
\begin{equation}
    F_{\mathbf{w}}: \left(\eta,\rho,v_0\right) \mapsto \left(z_{\rm D},z_{\rm C},z_{\rm O}\right),
\end{equation}
where $z_{\rm D}$, $z_{\rm C}$, and $z_{\rm O}$ are logits for the disordered, coexistence-candidate, and ordered labels. The logits are converted to classifier phase-assignment probabilities using the softmax map
\begin{equation}
    p_c(\boldsymbol{\lambda}) = \frac{e^{z_c(\boldsymbol{\lambda})}}{\sum_{c'}e^{z_{c'}(\boldsymbol{\lambda})}},
    \label{eq:softmax_probabilities}
\end{equation}
where $c$ runs over the three operational classes. These probabilities are
used to interpolate the finite-size phase map and to measure classifier
uncertainty. The network architecture is $3 \longrightarrow 128 \longrightarrow 3$ with a ReLU activation after the hidden layer and dropout with probability $0.2$~\cite{srivastava2014dropout}. The model is trained for $200$ epochs using the Adam optimizer with learning rate $\alpha = 10^{-3}$~\cite{kingma2014adam}. To compensate for class imbalance, we use a weighted cross-entropy loss
\begin{equation}
    \mathcal{L}(\mathbf{w}) = -\frac{1}{M}\sum_{i=1}^{M}w_{y_i}\log p_{y_i}(\boldsymbol{\lambda}_i),
    \label{eq:weighted_cross_entropy}
\end{equation}
where $M$ is the number of training points, $y_i$ is the operational label of $\boldsymbol{\lambda}_i$, and
\begin{equation}
    w_c = \frac{M}{3N_c}
\end{equation}
is the inverse-frequency weight for class $c$, with $N_c$ the number of
training examples in that class.

After training the classifier on the current data set, we generate a pool of $2000$ candidate parameter points. For each candidate point, we compute the Shannon entropy of the softmax probabilities~\cite{shannon1948mathematical}:
\begin{equation}
    H(\boldsymbol{\lambda}) = -\sum_c p_c(\boldsymbol{\lambda})\log_2 p_c(\boldsymbol{\lambda}).
    \label{eq:shannon_entropy}
\end{equation}
The entropy is small when the classifier assigns a point confidently to one class and large when two or more phase-assignment probabilities are comparable. High-entropy points are therefore expected to lie near the learned crossover surfaces. At each active-learning iteration, we select the $30$ candidate points with largest $H(\boldsymbol{\lambda})$, simulate them, assign operational labels using Eqs.~\eqref{eq:label_disorder}--\eqref{eq:label_order}, and add them to the training set. We perform seven active-learning iterations, producing a final data set of 310 simulated parameter points. The classifier is then retrained on the full data set and used to construct the finite-size phase map discussed in Sec.~\ref{sec:phase_diagram}.

The resulting active-learning data set is summarized in Fig.~\ref{fig:al_dataset}. Figure~\ref{fig:al_dataset}(a) shows the simulated points in observable space $\left(\left\langle\Phi\right\rangle,G,\log_{10}\chi\right)$, colored by the operational labels in Eqs.~\eqref{eq:label_disorder}--\eqref{eq:label_order}. Figure~\ref{fig:al_dataset}(b) shows the same points in parameter space $\left(\eta,\rho,v_0\right)$. The observable-space representation shows the separation between low-order, intermediate, and strongly ordered states, while the parameter-space representation shows how active learning populates the crossover regions between the operational classes.

The output of the active-learning procedure is therefore a finite-size classifier map over $\left(\eta,\rho,v_0\right)$, trained on reproducible global-order labels. This map is then used to visualize the learned crossover geometry and to select representative regions for direct physical validation.

\section{Finite-Size Phase Diagram}
\label{sec:phase_diagram}

We now use the classifier trained in Sec.~\ref{sec:active_learning} to interpolate the operational labels throughout the control-parameter space $\left(\eta,\rho,v_0\right)$. The resulting map is a finite-size phase diagram that resolves the morphology of the simulated $L=128$ system and provides crossover surfaces between operational regimes. This distinction is important because the Vicsek transition is known to involve strong finite-size effects and banded coexistence over broad parameter windows~\cite{chate2008collective}.

\begin{figure}[tbp!]
    \centering
    \includegraphics[width=1\linewidth]{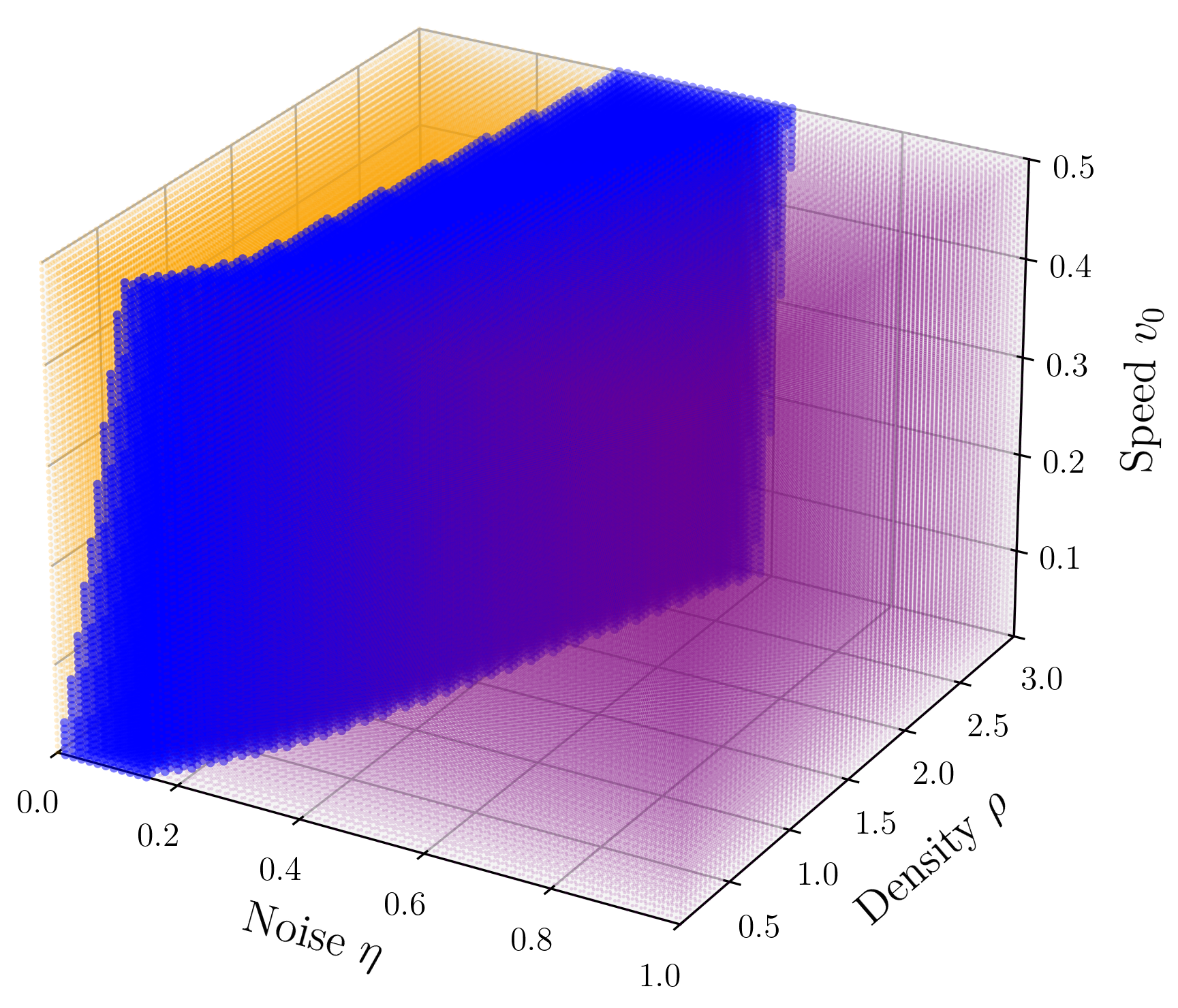}
    \caption{Three-dimensional classifier phase map in the control-parameter space $\left(\eta,\rho,v_0\right)$. The plotted points show the predicted operational class on a dense grid, with disordered, ordered, and coexistence-candidate regions shown in purple, orange, and blue, respectively. The coexistence-candidate region forms a curved sheet separating the high-noise disordered gas from the low-noise polar ordered regime.}
    \label{fig:phase_volume}
\end{figure}

Figure~\ref{fig:phase_volume} shows the learned classifier map in the full three-dimensional parameter space. The disordered gas occupies the large-noise region, where angular noise overwhelms local alignment. The polar ordered regime occupies the low-noise region, where alignment interactions produce a macroscopic flocking direction. Between them, the classifier identifies an intermediate sheet-like region. This region is the coexistence-candidate class defined in Sec.~\ref{sec:active_learning}, which is intermediate in global polar order, but its interpretation as banded coexistence is not assumed at this stage.

\begin{figure*}[tbp!]
    \centering
    \includegraphics[width=0.9\linewidth]{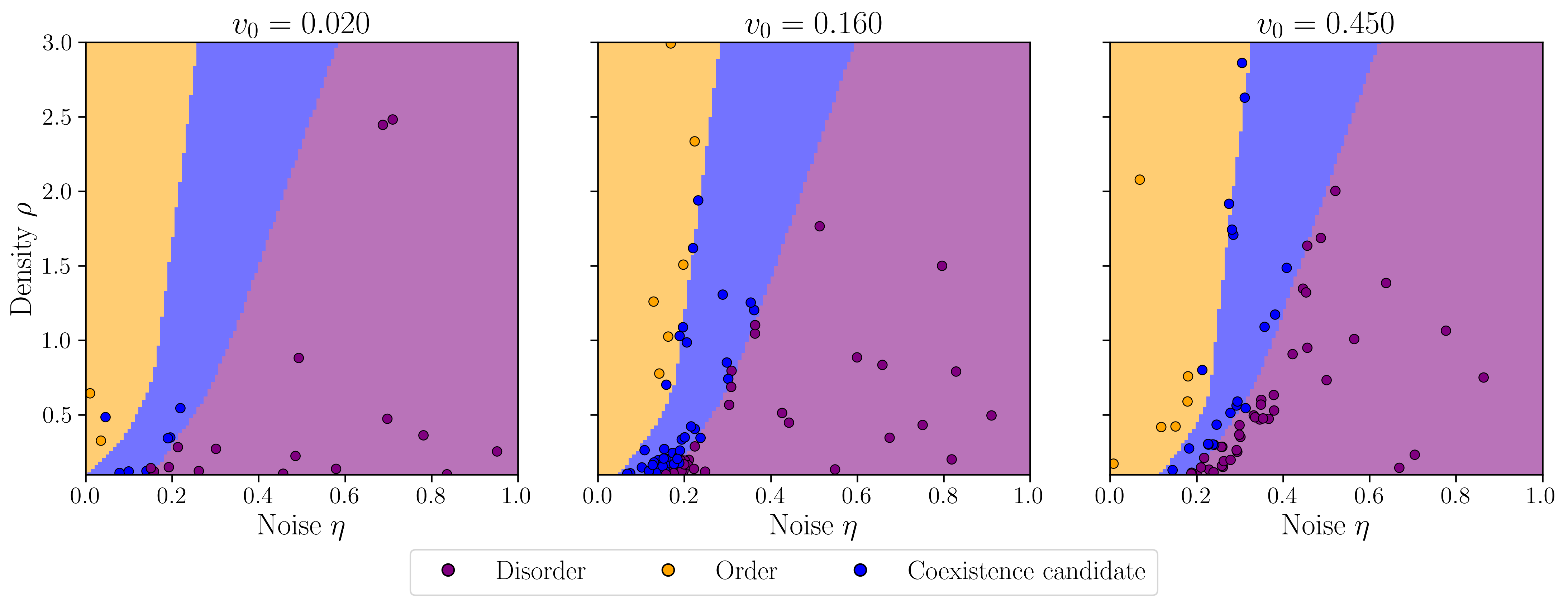}
    \caption{Two-dimensional slices of the learned finite-size phase diagram at fixed speed. The background color gives the classifier-predicted operational regime, while the overlaid points show simulated parameter configurations from the active-learning data set. From left to right, the slices show $v_0=0.020$, $v_0=0.160$, and $v_0=0.450$.}
    \label{fig:phase_slices}
\end{figure*}

The geometry of the learned regimes is clearer in two-dimensional slices at fixed speed. Figure~\ref{fig:phase_slices} shows slices through the classifier map at $v_0=0.020$, $v_0=0.160$, and $v_0=0.450$. In each slice, increasing noise drives the system from ordered to intermediate and then to disordered behavior. Increasing density shifts the crossovers to larger noise because each particle has more neighbors within the interaction radius, strengthening the local alignment field. This produces upward-curving crossover lines in the $\left(\eta,\rho\right)$ plane. The speed dependence reflects the competition between local alignment, self-propelled transport, and density modulation. At very small $v_0$, particles move slowly relative to the alignment update, so locally ordered structures can persist, but transport is weak. At larger $v_0$, particles sample space more rapidly and encounter more neighbors over a fixed number of steps. This increases the effective mixing and can stabilize global polar order to larger noise. At the same time, stronger transport alters the stability of localized high-density bands. In the slices shown in Fig.~\ref{fig:phase_slices}, the coexistence-candidate window remains between order and disorder but shifts and changes width as $v_0$ is increased. This is consistent with the physical picture that banded coexistence is controlled both by local alignment strength and the ability of density inhomogeneities to persist under self-propelled motion.

\begin{figure*}[tbp!]
    \centering
    \includegraphics[width=0.95\linewidth]{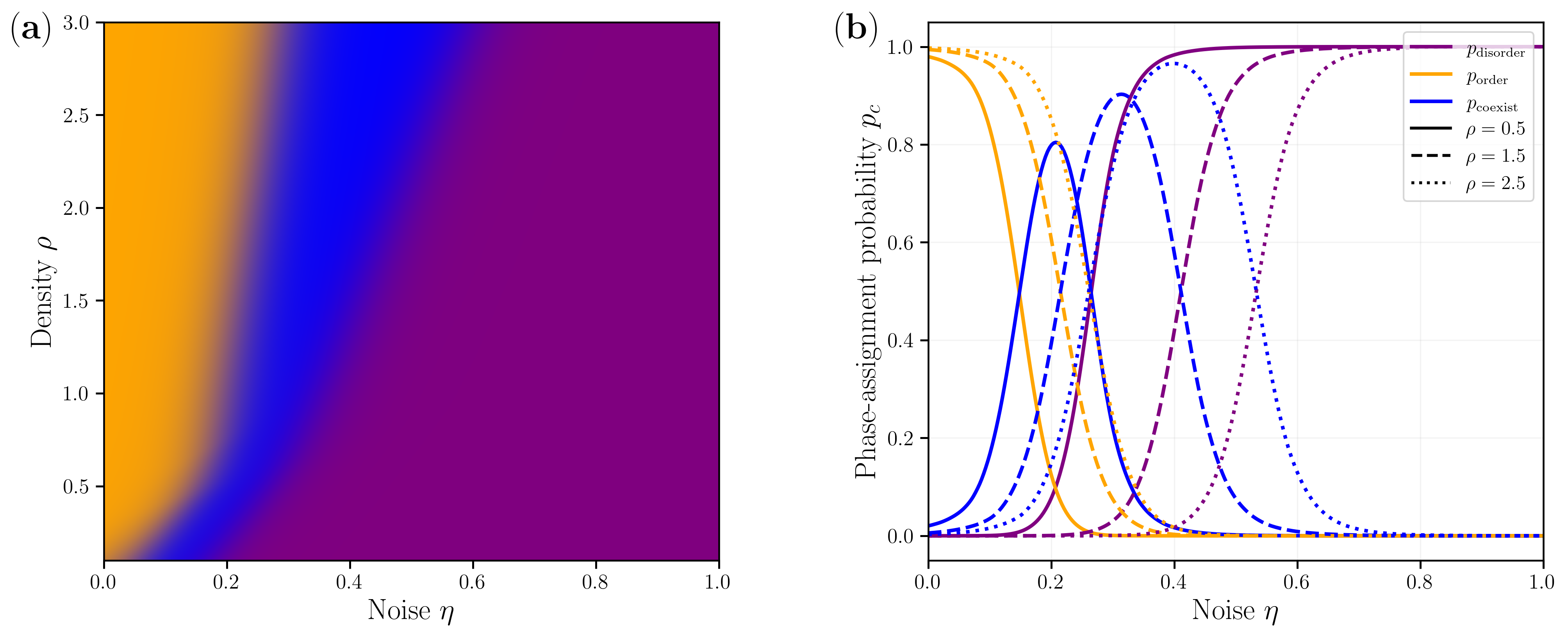}
    \caption{Soft classifier phase-assignment probabilities on the $v_0=0.16$ slice. (a)~Continuous probability map in the $\left(\eta,\rho\right)$ plane, obtained by mixing the disordered, ordered, and coexistence-candidate probabilities. (b) Probability cuts as functions of $\eta$ at fixed densities $\rho=0.5$, $\rho=1.5$, and $\rho=2.5$. Increasing density shifts the ordered-coexistence and coexistence-disorder crossovers to larger noise, reflecting the stronger local alignment field at higher particle density.}
    \label{fig:phase_probabilities}
\end{figure*}

This three-regime structure is consistent with the known phenomenology of the Vicsek model. The original study interpreted the onset of collective motion as a continuous transition~\cite{vicsek1995novel}, while later numerical work showed that the angular-noise model exhibits strong finite-size effects, discontinuous signatures, and propagating high-density ordered bands near onset~\cite{gregoire2004onset,chate2008collective}. Subsequent work connected this banded regime to a nonequilibrium liquid-gas or microphase-separation picture, in which dense ordered bands coexist with a dilute disordered background~\cite{solon2015phase,ginelli2016physics}. Hydrodynamic theories of polar active matter further emphasize that density and orientation fluctuations are strongly coupled, so that polar order can be accompanied by large density fluctuations and spatial inhomogeneity~\cite{toner1995long,toner1998flocks}. The learned phase map reproduces the corresponding qualitative organization: a high-noise gas, a low-noise polar ordered regime, and an intervening finite-width coexistence-candidate region.

Another observation from Fig.~\ref{fig:phase_slices} is that the coexistence-candidate region occupies a density-dependent noise interval. Near the bottom of the $\left(\eta,\rho\right)$ slices, the blue region remains present but is relatively narrow in $\eta$. At larger densities, the same region extends over a broader noise range and shifts to larger $\eta$. The learned map then suggests that the intermediate global-order regime becomes more robust as density is increased, which is discussed in more depth below.

To examine the finite-size crossover more continuously, we evaluate the softmax probabilities defined in Eq.~\eqref{eq:softmax_probabilities} on the $v_0=0.16$ slice. Figure~\ref{fig:phase_probabilities}(a) shows the resulting classifier phase-assignment probabilities as a color mixture over $\left(\eta,\rho\right)$, while Fig.~\ref{fig:phase_probabilities}(b) shows one-dimensional cuts at representative densities. These curves make explicit that the dominant phase-assignment probability changes smoothly across a finite noise interval, arising from both the underlying finite-size rounding of the simulated system and the continuous interpolation performed by the neural network.

\begin{figure}[tbp!]
    \centering
    \includegraphics[width=0.9\linewidth]{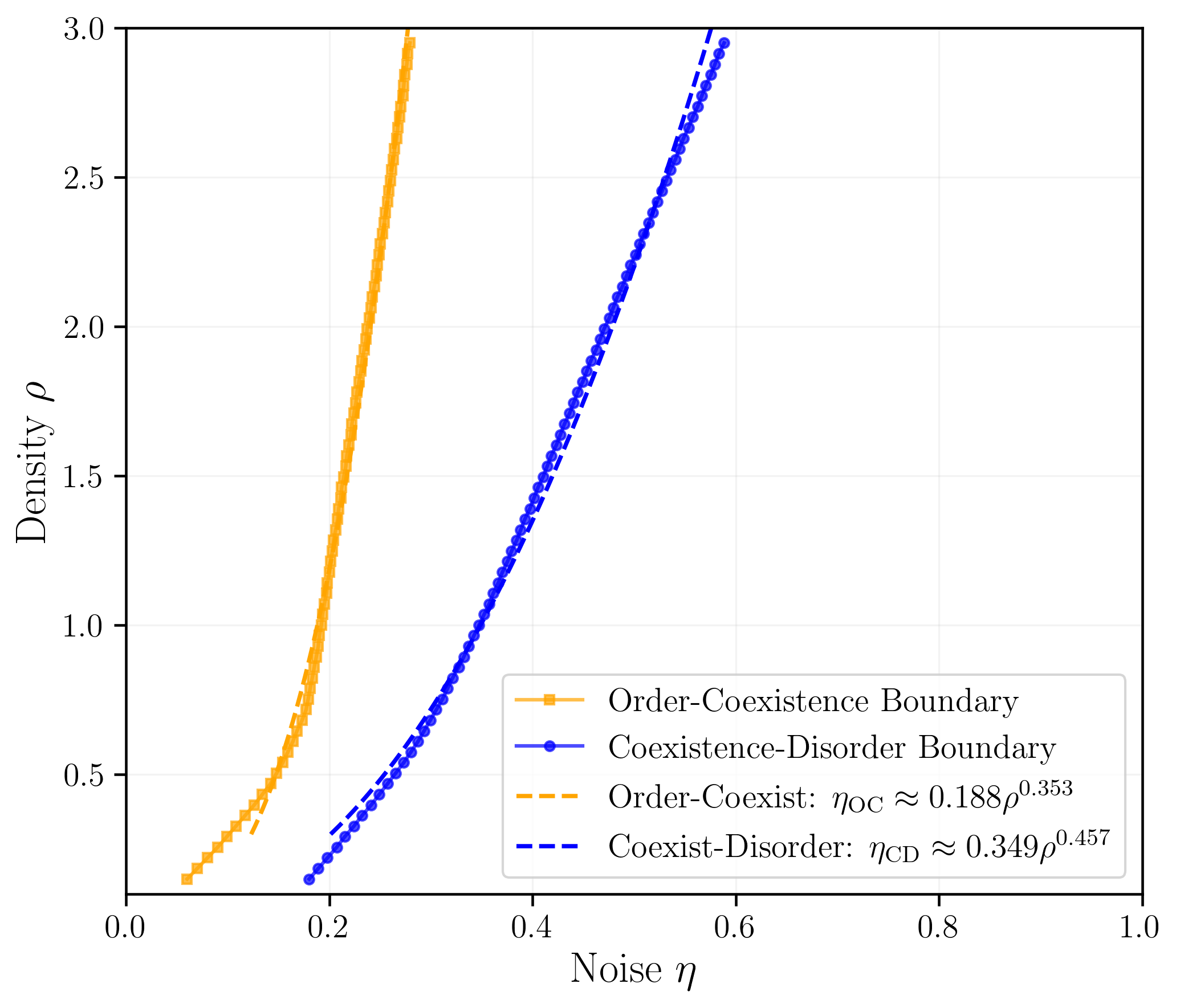}
    \caption{Effective crossover curves extracted from the learned classifier map at fixed speed $v_0=0.16$. The orange curve gives the ordered-to-coexistence crossover $\eta_{\rm OC}(\rho)$, while the blue curve gives the coexistence-to-disorder crossover $\eta_{\rm CD}(\rho)$. Dashed lines show power-law fits over the sampled density range. The coexistence-candidate window corresponds to the noise interval between the two curves and broadens with density.}
    \label{fig:boundary_curves}
\end{figure}

To quantify the density dependence of the learned crossovers, we extract two effective boundary curves from the $v_0=0.16$ classifier slice. The lower-noise curve $\eta_{\rm OC}(\rho)$ marks the crossover from the polar ordered regime to the coexistence-candidate regime as noise is increased. The higher-noise curve $\eta_{\rm CD}(\rho)$ marks the crossover from the coexistence-candidate regime to the disordered gas. Both curves increase with density, reflecting the stabilizing effect of larger local particle number on polar alignment. The extracted curves are well described over the sampled density range by power laws:
\begin{align}
    \eta_{\rm OC}(\rho) &\simeq 0.188\,\rho^{0.353},
    \label{eq:oc_boundary_fit} \\
    \eta_{\rm CD}(\rho) &\simeq 0.349\,\rho^{0.457}.
    \label{eq:cd_boundary_fit}
\end{align}
At the validation density $\rho_0=2.0$, these fits give $\eta_{\rm OC}\simeq 0.241$ and $\eta_{\rm CD}\simeq 0.479$, consistent with the ordered, coexistence-candidate, and disordered regions used in the spatial diagnostics. We interpret the exponents in Eqs.~\eqref{eq:oc_boundary_fit} and \eqref{eq:cd_boundary_fit} as effective finite-size crossover exponents rather than universal critical exponents, since they are obtained at a single system size and from classifier-defined operational boundaries. Their numerical values are very close to the exponents reported in the original Vicsek study, but they do not have the same interpretation. The original values $0.35$ and $0.45$ describe scaling of the order parameter near the onset of collective motion~\cite{vicsek1995novel}, whereas the present exponents, $\beta_{\rm OC}\simeq0.353$ and $\beta_{\rm CD}\simeq0.457$, describe how classifier-defined finite-size crossover boundaries shift with density. Later numerical work has also emphasized that the apparent transition location and scaling behavior depend on finite-size effects, density, and particle speed~\cite{baglietto2012criticality, rubio2019self, ginelli2016physics}. We therefore use these exponent comparisons only as consistency checks. 

The main robust result is that both classifier-defined boundaries shift to larger noise with increasing density. Moreover, because the higher-noise coexistence-to-disorder boundary grows slightly faster than the lower-noise ordered-to-coexistence boundary, the coexistence-candidate interval
\begin{equation}
    \Delta\eta_{\rm C}(\rho) = \eta_{\rm CD}(\rho)-\eta_{\rm OC}(\rho),
\end{equation}
widens over the sampled density range. This widening is directly visible in Fig.~\ref{fig:boundary_curves}. Physically, increasing density raises the typical number of neighbors inside the interaction radius, strengthening local alignment and allowing partially ordered, spatially inhomogeneous states to persist over a wider range of angular noise. The spatial content of this intermediate window is tested directly in Sec.~\ref{sec:spatial_validation}.

\begin{figure}[bp!]
    \centering
    \includegraphics[width=\linewidth]{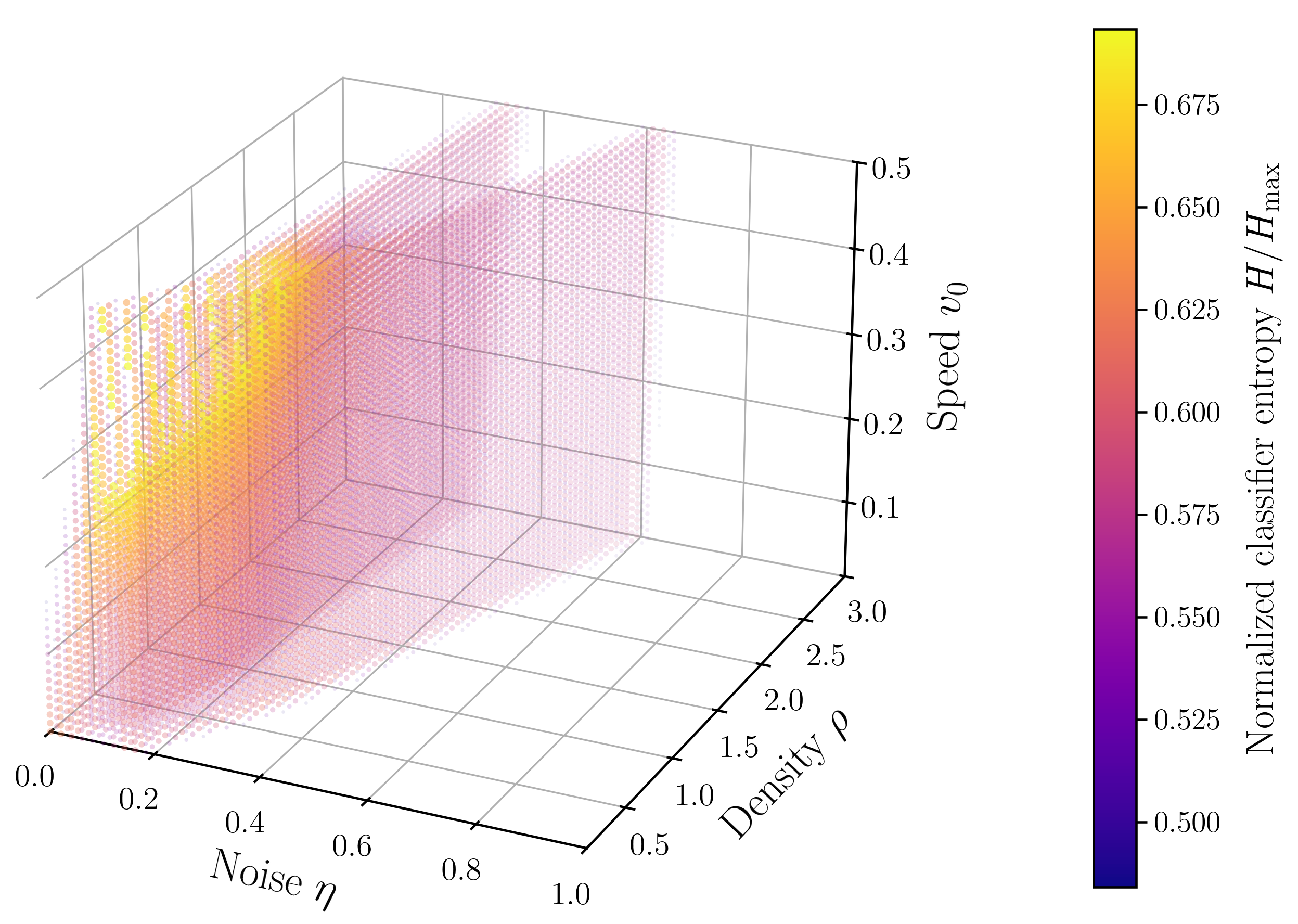}
    \caption{Normalized classifier entropy field in the three-dimensional parameter space $\left(\eta,\rho,v_0\right)$. Bright regions indicate large Shannon entropy of the softmax probabilities and therefore high classifier uncertainty. The high-entropy points form sheets near the learned crossover surfaces, showing that entropy acquisition targets the interfaces between operational regimes.}
    \label{fig:entropy_sheets}
\end{figure}

Finally, the entropy field used during active learning gives another complementary view of the learned crossover structure. Figure~\ref{fig:entropy_sheets} shows the normalized Shannon entropy $H/H_{\max}$, with $H_{\max}=\log_2 3$, computed from the softmax phase-assignment probabilities. We note that this quantity isn't a thermodynamic entropy; it measures only the ambiguity of the classifier assignment. The dominant high-entropy structures form sheet-like regions in $\left(\eta,\rho,v_0\right)$ space, aligned with the crossover surfaces visible in Figs.~\ref{fig:phase_volume}--\ref{fig:phase_probabilities} along the two sides of the coexistence-candidate regime. The entropy is especially elevated in parts of the low-density region, where local neighborhoods are sparse and the operational classes are less sharply separated at finite size. Away from these crossover and dilute uncertain regions, the classifier assigns points more confidently to a single operational class. The entropy field therefore provides an internal consistency check that the active-learning rule concentrates new simulations near ambiguous regions of the finite-size phase map rather than deep inside well-resolved ordered, disordered, or coexistence-candidate regimes.

\begin{figure*}[tbp!]
    \centering
    \includegraphics[width=\linewidth]{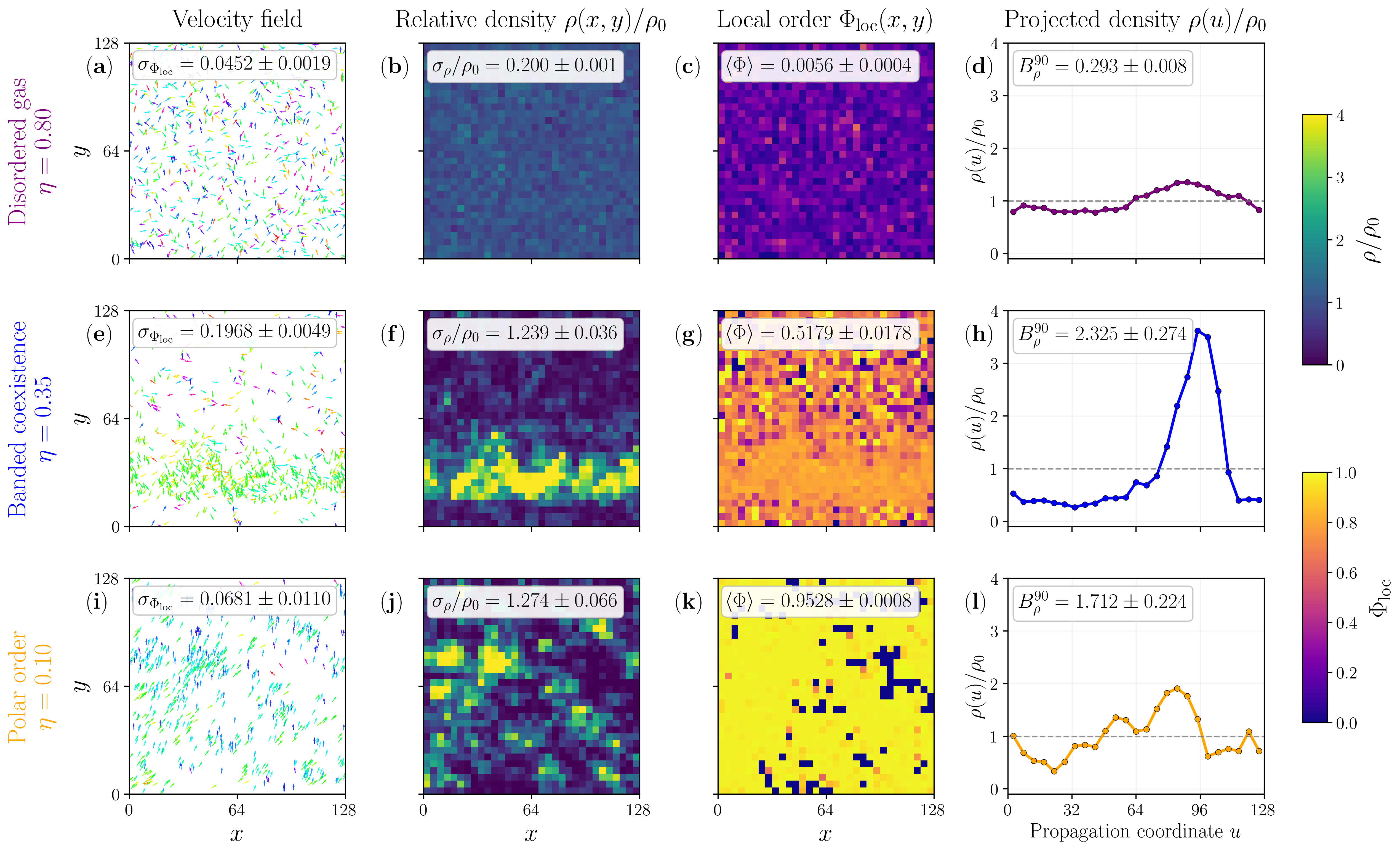}
    \caption{Spatial validation of the three operational regimes at fixed $\rho_0=2.0$ and $v_0=0.16$. Rows show a high-noise disordered gas $(\eta=0.80)$, an intermediate banded coexistence state $(\eta=0.35)$, and a low-noise polar ordered state $(\eta=0.10)$. Columns show the velocity field (with arrows colored by particle heading), the relative density field $\rho(x,y)/\rho_0$, the local polar-order field $\Phi_{\rm loc}(x,y)$, and the projected density profile $\rho(u)/\rho_0$ along the propagation coordinate. Boxed values are averages over five independent trials. The disordered state is spatially diffuse and weakly ordered, the intermediate state displays a dense, locally ordered band, and the ordered state remains locally ordered across most of the system despite strong density fluctuations.}
    \label{fig:physical_validation}
\end{figure*}

\section{Spatial Diagnostics and Validation of Banded Coexistence}
\label{sec:spatial_validation}

The classifier map identifies an intermediate operational regime using only the global order parameter $\left\langle\Phi\right\rangle$. However, intermediate global order is not a unique morphological signature. In Vicsek-type flocking models, such a value is consistent with a microphase-separated state in which dense ordered bands travel through a dilute disordered background~\cite{solon2015phase,chate2020dry}. However, it may also reflect finite-size rounding near a crossover, partial global alignment without a sharply resolved band, or density inhomogeneities within an otherwise polar ordered flock, since density and orientation fluctuations remain strongly coupled in the ordered phase~\cite{toner1998flocks, mahault2019quantitative}. We therefore validate the intermediate regime using spatially resolved diagnostics that were not used to train the classifier.

We focus on the fixed density $\rho_0 = 2.0$ and speed $v_0 = 0.16$ for which the classifier predicts an ordered regime at low noise, a coexistence-candidate regime at intermediate noise, and a disordered gas at high noise. We compare representative noise values $\eta = 0.10, 0.35, 0.80$, corresponding respectively to the polar ordered, coexistence-candidate, and disordered regimes. For each noise value, the diagnostics reported below are averaged over five independent trials, and the plotted spatial fields show one representative final configuration.

To quantify spatial structure, we coarse grain the simulation domain into square bins. Let $\mathcal{C}_{mn}$ denote one such bin, $\mathbf{x}_{mn}$ denote its center, and $n_{mn}$ be the number of particles inside it. If the bin area is $a^2$, the coarse-grained density is
\begin{equation}
    \rho_{mn} = \frac{n_{mn}}{a^2}.
\end{equation}
The corresponding local polar order is
\begin{equation}
    \Phi_{\rm loc}(\mathbf{x}_{mn}) = \frac{1}{n_{mn}}\left|\sum_{j\in\mathcal{C}_{mn}} e^{i\theta_j}\right|
    \label{eq:local_order_field}
\end{equation}
for occupied bins. We use the spatial standard deviations $\sigma_\rho/\rho_0$ and $\sigma_{\Phi_{\rm loc}}$ to measure density inhomogeneity and local-order heterogeneity. We also compute the density-order correlation
\begin{equation}
    C_{\rho,\Phi} = \operatorname{corr}\left(\rho_{mn}, \Phi_{\rm loc}(\mathbf{x}_{mn})\right),
    \label{eq:density_order_correlation}
\end{equation}
which distinguishes dense ordered bands from density fluctuations that are only weakly tied to local alignment. Finally, to measure band contrast along the direction of collective motion, we project particle positions onto the instantaneous flocking direction
\begin{equation}
    \hat{\mathbf{p}} = \frac{\sum_j \mathbf{e}_j}{\left|\sum_j \mathbf{e}_j\right|}
\end{equation}
and define the propagation coordinate
\begin{equation}
    u_i = \mathbf{r}_i\cdot\hat{\mathbf{p}} \pmod L.
\end{equation}
The projected density profile $\rho(u)$ is obtained by binning particles along $u$. We define a robust band contrast
\begin{equation}
    B_\rho^{90} = \frac{P_{95}\!\left[\rho(u)\right] - P_{5}\!\left[\rho(u)\right]}{\rho_0},
    \label{eq:band_contrast}
\end{equation}
where $P_{95}$ and $P_5$ are the 95th and 5th percentiles of the projected
density profile. This percentile-based contrast is less sensitive to isolated
empty or overfilled bins than a maximum-minus-minimum contrast.

\begin{figure*}[tbp!]
    \centering
    \includegraphics[width=\linewidth]{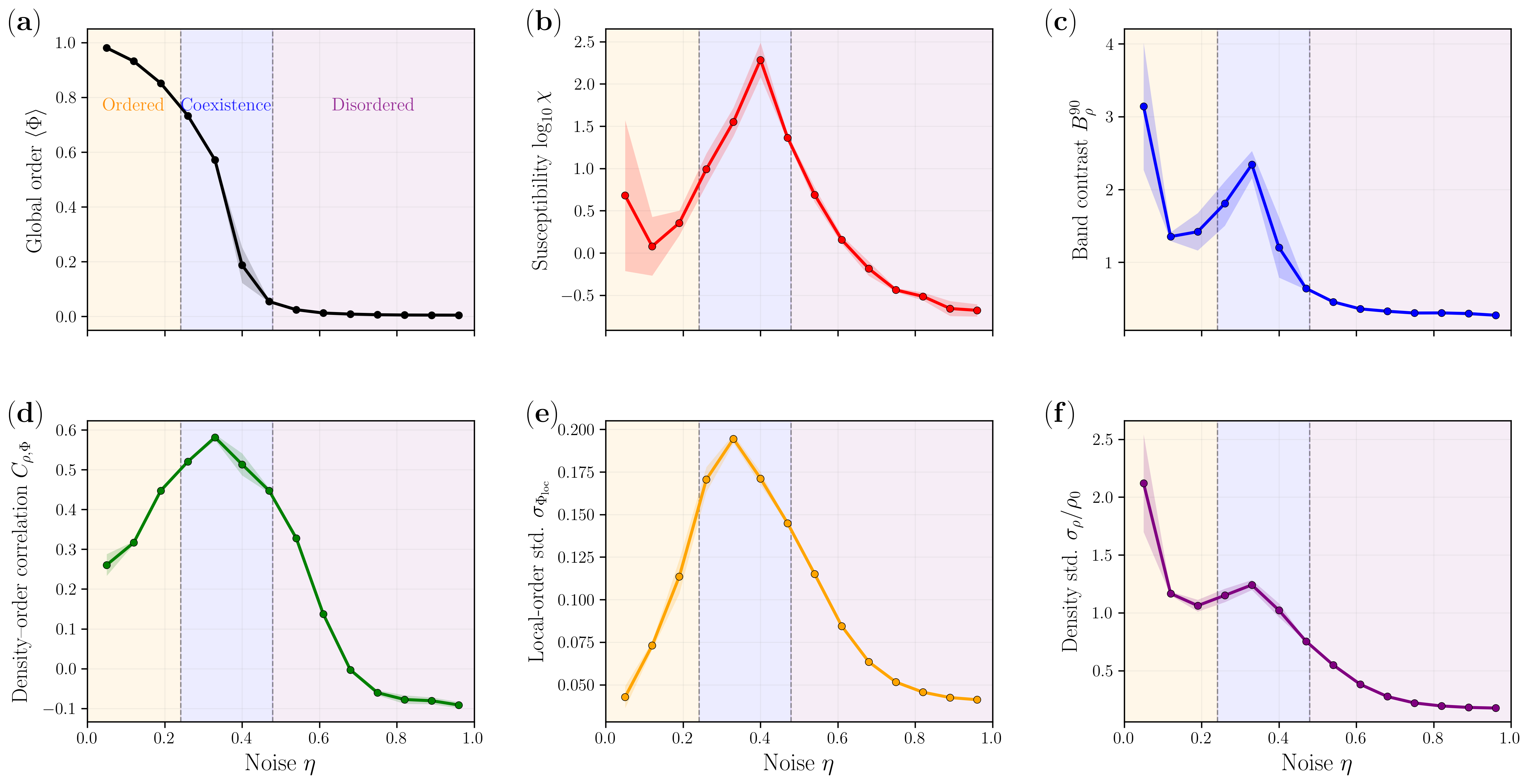}
    \caption{Fixed-density noise scan at $\rho_0=2.0$ and $v_0=0.16$. Shaded regions indicate the ordered, coexistence-candidate, and disordered intervals predicted by the classifier. Curves show averages over five independent trials, with shaded bands indicating trial-to-trial variation. The scan compares (a)~global polar order, (b)~susceptibility, (c)~projected band contrast, (d)~density-order correlation, (e)~spatial variation of local polar order, and (f)~relative density fluctuations. The coexistence window is marked by the simultaneous enhancement of several of these spatial diagnostics.}
    \label{fig:noise_scan}
\end{figure*}

Figure~\ref{fig:physical_validation} compares the spatial structure of the three representative regimes. In the disordered gas, the velocity field is randomly oriented, the density field is nearly homogeneous, and the local polar-order field remains weak throughout the domain. Quantitatively, the global order parameter $\left\langle\Phi\right\rangle = 0.0056\pm0.0004$ is close to the finite-size noise floor, and the projected density contrast $B_\rho^{90} = 0.2928\pm0.0076$ is small. The local-order variation $\sigma_{\Phi_{\rm loc}} = 0.0452\pm0.0019$ is also weak, and the density-order correlation $C_{\rho,\Phi} = -0.0725\pm0.0063$ is slightly negative. Thus, beyond being just globally disordered, the disordered state is also spatially diffuse and lacks any systematic association between high density and local alignment.

The intermediate state at $\eta=0.35$ has a qualitatively different spatial structure. The velocity snapshot shows a coherent band crossing the system, and the density map contains a dense strip embedded in a more dilute background. The projected density profile exhibits a sharp peak reaching several times the mean density, giving $B_\rho^{90} = 2.3248\pm0.2745$. At the same time, the local polar-order field is strongly heterogeneous, since regions inside the dense band are highly ordered, while the surrounding dilute gas is much less ordered. This produces by far the largest local-order variation of the three representative states: $\sigma_{\Phi_{\rm loc}} = 0.1968\pm0.0049$. The density-order correlation $C_{\rho,\Phi} = 0.5886\pm0.0125$ is also maximal, showing that the high-density region is a locally aligned structure, not just a density fluctuation. The global order parameter $\left\langle\Phi\right\rangle = 0.5179\pm0.0178$ remains intermediate because the system contains both an ordered dense band and a less ordered background. This is the spatial signature expected for banded coexistence.

The ordered state at $\eta=0.10$ provides an important control case. It has very large global polar order $\left\langle\Phi\right\rangle = 0.9528\pm0.0008$, and the velocity field is aligned across nearly the entire system. However, the density field is still strongly inhomogeneous: $\sigma_\rho/\rho_0 = 1.2737\pm0.0656$, which is comparable to the coexistence value $\sigma_\rho/\rho_0 = 1.2390\pm0.0355$. Density fluctuations alone are therefore not sufficient to identify banded coexistence. The distinction is instead that the ordered state remains locally ordered almost everywhere, giving a much smaller local-order variation $\sigma_{\Phi_{\rm loc}} = 0.0681\pm0.0110$ and a weaker density-order correlation $C_{\rho,\Phi} = 0.2609\pm0.0019$. The ordered phase can therefore exhibit substantial density inhomogeneity, as expected from density-orientation coupling in polar active matter, without showing the coexistence morphology of a dense ordered band traveling through a dilute, weakly ordered background.

The fixed-density noise scan in Fig.~\ref{fig:noise_scan} shows that the same diagnostics vary systematically across the classifier-predicted regimes. The global order parameter in Fig.~\ref{fig:noise_scan}(a) decreases monotonically with increasing noise, passing from a nearly ordered flock to a disordered gas. This confirms that the operational labels based on $\left\langle\Phi\right\rangle$ capture the overall ordering trend. However, the additional panels show that the intermediate regime has a distinct spatial signature rather than merely an intermediate value of global order.

The susceptibility in Fig.~\ref{fig:noise_scan}(b) peaks inside the coexistence-candidate interval, indicating enhanced fluctuations of the global order parameter near the banded regime. The projected band contrast in Fig.~\ref{fig:noise_scan}(c) is also large in and near the coexistence window, with a pronounced maximum around the representative banded point. However, the contrast does not vanish in the ordered regime, which again shows that density inhomogeneity by itself is not a unique identifier of coexistence. Instead, the most selective signatures are the density-order correlation and the local-order heterogeneity. Figure~\ref{fig:noise_scan}(d) shows that $C_{\rho,\Phi}$ rises through the ordered-to-coexistence crossover and reaches its largest value in the intermediate regime, meaning that dense regions are also the most locally aligned there. Figure~\ref{fig:noise_scan}(e) shows a similar peak in $\sigma_{\Phi_{\rm loc}}$, indicating the coexistence of locally ordered and locally disordered regions within the same finite system. Both quantities fall at high noise, where the system becomes a disordered gas.

The relative density fluctuation in Fig.~\ref{fig:noise_scan}(f) provides a useful caution. It is large at low noise and remains sizable through the coexistence interval, then decreases as the system enters the high-noise gas. This behavior is consistent with the giant density fluctuations and strong inhomogeneities expected in polar active matter, showing why a density-only criterion would be ambiguous, namely, because the ordered state can have large density fluctuations while remaining locally aligned throughout the system. Banded coexistence is instead identified by the joint occurrence of a dense projected band, strong local-order heterogeneity, and a large positive correlation between density and local order.

Taken together, Figs.~\ref{fig:physical_validation} and \ref{fig:noise_scan} validate the physical interpretation of the intermediate classifier regime along the fixed-density scan. Although the classifier identifies this region only through its intermediate global polar order, the spatial diagnostics show that it contains dense, locally ordered bands coexisting with a dilute, weakly ordered background. The ordered regime provides the complementary control case, showing that strong density fluctuations can occur without destroying local alignment across most of the system.

\section{Discussion and Conclusions}
\label{sec:conclusion}

We have used active learning to construct a finite-size phase map of the two-dimensional angular-noise Vicsek model in the control-parameter space $\left(\eta,\rho,v_0\right)$. The learned map identifies three operational regimes: a high-noise disordered gas, a low-noise polar ordered regime, and an intermediate coexistence-candidate regime. At fixed $v_0$, the extracted crossover curves shift to larger noise with increasing density, consistent with the physical expectation that larger local particle number strengthens alignment and stabilizes collective motion.

The main methodological point is that the classifier is a tool for adaptive sampling and interpolation, not a replacement for physical diagnostics. Because it is trained only on labels derived from $\left\langle\Phi\right\rangle$, it can locate where intermediate global order occurs but cannot determine the spatial morphology of that state. The spatial validation shows that this distinction matters: the coexistence-candidate regime contains a dense, locally ordered band, while the low-noise ordered regime can exhibit density fluctuations of comparable magnitude without losing local alignment across most of the system. Therefore, the important diagnostic outcome is whether density concentration is accompanied by strong local-order heterogeneity and a large positive density-order correlation.

These results also clarify the role of machine learning in the analysis. Although the boundary fits provide useful summaries of the learned map, the main value of the neural network is to act as an adaptive surrogate for an expensive simulation space. It identifies uncertain crossover regions, guides the placement of new simulations, produces a smooth representation of the finite-size phase geometry, and selects representative regions for direct physical validation. The subsequent boundary fits summarize the learned map, but the scientific interpretation comes from combining the classifier with independent spatial measurements.

Several extensions follow naturally from this framework. A systematic finite-size scaling study would be needed to determine which features of the learned crossover surfaces persist in the thermodynamic limit, while applying the same spatial diagnostics across the full coexistence-candidate sheet would clarify how band morphology varies throughout $\left(\eta,\rho,v_0\right)$. A future paper will study hysteresis and protocol dependence near the Vicsek transition~\cite{baihysteresis}, in which we will compare phase-assignment vectors obtained under different preparation histories, such as increasing-noise and decreasing-noise sweeps. This would move beyond the static crossover geometry studied here and connect the learned finite-size phase map to metastability and history dependence in nonequilibrium flocking transitions.

Overall, this work demonstrates a workflow for studying nonequilibrium phase structure in active-matter simulations. Active learning is used to locate and resolve finite-size crossover regions efficiently, while independent spatial diagnostics are used to assign physical meaning to the learned regimes. Applied to the Vicsek model, this approach recovers the expected organization into disordered, banded coexistence, and polar ordered morphologies, and it illustrates how machine-learning phase maps can be converted from operational classifications to physically interpretable descriptions of nonequilibrium states.

\bibliography{refs}
%\nocite{*}
\end{document}